\documentclass[preprint,showpacs,preprintnumbers,amsmath,amssymb]{revtex4}

\usepackage{graphicx}% Include figure files
\usepackage{dcolumn}% Align table columns on decimal point
\usepackage{bm}% bold math

\begin{document}

\title{Thermal Conductivity of Anharmonic Lattices:\\
Effective Phonons and Quantum Corrections}

\author{Dahai He$^{1}$} \email{dhhe@hkbu.edu.hk}
\author{Sahin Buyukdagli$^{1}$} \email{sbuyukda@hkbu.edu.hk}

\author{Bambi Hu$^{1,2}$}
\affiliation{%
$^{1}$Department of Physics, Centre for Nonlinear Studies, and the
Beijing- Hong Kong- Singapore Joint Centre for Nonlinear and
Complex Systems (Hong Kong), Hong Kong Baptist
University, Kowloon Tong, Hong Kong, China\\
$^{2}$Department of Physics, University of Houston, Houston, Texas
77204-5005, USA
}%

\begin{abstract}
We compare two effective phonon theories, which have both been
applied recently to study heat conduction in anharmonic lattices.
In particular, we study the temperature dependence of the thermal
conductivity of Fermi-Pasta-Ulam $\beta$ model via the Debye formula,
showing the equivalence of both approaches. The temperature for
the minimum of the thermal conductivity and the corresponding
scaling behavior are analytically calculated, which agree well
with the result obtained from non-equilibrium simulations. We also
give quantum corrections for the thermal conductivity from quantum
self-consistent phonon theory. The vanishing behavior at low
temperature regime and the existence of an \textit{umklapp} peak
are qualitatively consistent with experimental studies.
\end{abstract}

\date{\today}                            %display desired
\maketitle

The study of heat conduction is very important from both theoretical
and experimental point of view. A traditional phenomenological
approach to understand the thermal properties in solids is the
Debye formula given by
\begin{equation}\label{Debye}
\kappa=\sum_k{C_kv_kl_k},
\end{equation}
where $\kappa$ is the thermal conductivity, $C_k$, $v_k$, $l_k$
are the specific heat, the phonon group velocity and the phonon
mean free path of mode $k$, respectively. In spite of its
successfulness for qualitative explanation of heat conduction in
dielectrics, quantitative predictions are hard to make from a
microscopic viewpoint.

Recently, an increasing study of heat conduction in low
dimensional Hamiltonian models may shed light on its microscopic
understanding~\cite{Lepri_rep}. However, only a few integrable
models can be solved rigourously~\cite{Rieder67}. Generally one
has to rely on numerical simulations for non-integrable models.
Thus it would be worthy to revisit the traditional kinetic
approach by incorporating some microscopic consideration for the
low dimensional non-integrable lattice systems.

According to the Debye formula, the thermal transport process in a
nonlinear lattice is intrinsically relative to its dispersion
relation and relaxation of normal modes (phonons). The existence
of nonlinearity makes the definition of phonon delicate, and it is
even harder in this case to quantify  phonon transport from a
first-principle way. To surmount the difficulties due to
nonlinearity, the concept of ``effective phonons'' has been
recently introduced to study heat conduction in dynamical
models~\cite{Li_EPT06, Li_EPT07a, Li_EPT07b, Hu_SCPT06}. The basic
idea consists in incorporating the non-linearity into normal modes
by renormalizing the harmonic frequency spectrum. In the serial
studies~\cite{Li_EPT06, Li_EPT07a, Li_EPT07b}, the authors apply
the so-called effective phonon theory (EPT) to study heat
conduction within the kinetic framework. In Ref.~\cite{Hu_SCPT06},
the authors study heat conduction through a lattice consisting of
two weakly coupled nonlinear segments via the self-consistent
phonon theory (SCPT).

It is thus interesting to compare EPT and SCPT since they have
both been applied to the field of heat conduction. In the present
study, we will give a detailed comparison via the Debye formula as
done in~\cite{Li_EPT06, Li_EPT07a, Li_EPT07b}. Our result shows
the equivalence of SCPT and EPT in a large range of temperature.
Considering the failure of the classical description at low
temperature regime, we also compute quantum corrections to the thermal
conductivity by extending our study to quantum regime, which gives
qualitatively consistent results with experimental studies.

It should be emphasized that both EPT and SCPT cannot give the
microscopic definition of the relaxation time $\tau_k$ for phonons
of mode $k$. A traditional perturbative way consists in studying
the single mode relaxation time based on three- and four- phonon
processes~\cite{Srivastava90}. Great efforts have been devoted to
obtain the relaxation time of the heat current
correlations~\cite{Lepri98a, Lepri98b}, which in general might be
related to the relaxation time of phonons. However, the relation
was so far unclear. For an anharmonic chain, a simple but
physically appealing assumption
\begin{equation}\label{tau}
\tau_k^{-1}\propto{\omega_k\epsilon}
\end{equation}
has been proposed in Ref.~\cite{Li_EPT07a}, where $\omega_k$ is
the phono frequency of mode $k$. The dimensionless nonlinearity
$\epsilon$ is defined as the ratio between the average of the
anharmonic potential energy and the total potential energy:
\begin{equation}
\epsilon=\frac{|\langle E_n\rangle|}{\langle E_l+E_n \rangle}.
\end{equation}
One can see that, when $\epsilon$ vanishes, $\tau$ approaches to
infinity, leading to the expected divergence of the thermal
conductivity. In the following, we will apply the assumption of
Eq.~\eqref{tau} to Eq.~\eqref{Debye} in order to study the thermal
conduction of an anharmonic chain, which gives surprising
agreement with non-equilibrium simulations.

\section{Effective phonon approaches}
In this section we will compare the SCPT (see, e.g.,
\cite{Bruesch82, Dauxois93}) and EPT~\cite{Alabiso95, Alabiso01}.
To make this concrete, we mainly focus on the temperature
dependence of the physical quantities of the Fermi-Pasta-Ulam
$\beta$ (FPU-$\beta$) model, which is a classic example to study
heat conduction. The Hamiltonian of the FPU-$\beta$ model is given
by
\begin{equation}\label{Ham}
H=\sum
\frac{p_i^2}{2}+\frac{K}{2}(x_{i+1}-x_i)^2+\frac{\lambda}{4}(x_{i+1}-x_i)^4.
\end{equation}
Within the SCP approximation, the Hamiltonian~\eqref{Ham} is
approximated by a trial Hamiltonian
\begin{equation}\label{HamSCPT}
H_0=\sum \frac{p_i^2}{2}+\frac{f}{2}(x_{i+1}-x_i)^2,
\end{equation}
where the effective harmonic coupling constant $f$ is given by
Eq.~\eqref{f} in the appendix with the Boltzmann constant $k_B=1$.
Note that $f$ is temperature dependent, which stems from the
existence of nonlinearity in the system. The dispersion relation
of effective phonons corresponding to the effective
Hamiltonian~\eqref{HamSCPT} can be written in the form
\begin{equation}\label{DRSCPT}
\omega_k=2\sqrt{f}\sin{\frac{k}{2}}.
\end{equation}
It is then straightforward to calculate the dimensionless nonlinearity
$\epsilon$ and the specific heat $C$ from gaussian averages of SCPT, which yields
\begin{equation}\label{epsSCPT}
\epsilon=\frac{f-K}{f+K},
\end{equation}
\begin{equation}\label{SHSCPT}
C=\frac{3}{4}+\frac{K}{4\sqrt{K^2+12\lambda T}}.
\end{equation}

The derivation of the effective phonon spectrum is based on the
generalized equipartition theorem,
\begin{equation}\label{equip}
k_BT=\left<q_k\frac{\partial H}{\partial q_k}\right>_c,
\end{equation}
where the bracket $\left<\cdot\right>_c$ stands for a thermal
average in the canonical ensemble and $q_k$ denotes the Fourier
transform of the coordinate $x_i$. The next step consists of an
approximative transformation of the right hand side of
Eq.~\eqref{equip} into a more compact form
(see~\cite{Alabiso95,Li_EPT06} for details),
\begin{equation}\label{equip2}
k_BT=\tilde{\omega}_k^2\left<q_k^2\right>_c.
\end{equation}
Specifically, the effective phonon spectrum $\tilde{\omega}_k$ for
the FPU-$\beta$ model reads as
\begin{equation}\label{DREPT}
\tilde{\omega}_k=2\sqrt{\alpha}\sin{\frac{k}{2}}.
\end{equation}
Here $\alpha$ is given by
\begin{equation}\label{alpha}
\alpha=K+\lambda\frac{\langle \sum_i{(x_{i+1}-x_i)^4}\rangle_c}
{\langle \sum_i{(x_{i+1}-x_i)^2}\rangle_c} =\frac{ 2\lambda T
Y_{1/4}(x)}{K [Y_{3/4}(x)-Y_{1/4}(x)] },
\end{equation}
where $Y_{1/4}(x)$ and $Y_{3/4}(x)$ are the modified Bessel
functions of the second kind and $x\equiv K^2/(8\lambda T)$. The
parameters for the nonlinearity $\tilde{\epsilon}$ and the specific heat
$\tilde{C}$ that follow from EPT can be expressed in the form
\begin{equation}\label{epsEPT}
\tilde{\epsilon}=\frac{K^2[Y_{1/4}(x)-Y_{3/4}(x)]+2\lambda T
Y_{1/4}(x)} {K^2[Y_{3/4}(x)-Y_{1/4}(x)]+2\lambda T Y_{1/4}(x)},
\end{equation}
\begin{equation}\label{SHEPT}
\tilde{C}=\frac{3}{4}+\frac{K^2}{16\lambda
T}\frac{Y_{3/4}(x)}{Y_{1/4}(x)}+\frac{K^4}{64\lambda^2T^2}
(1-\frac{Y_{3/4}(x)^2}{Y_{1/4}(x)^2}).
\end{equation}

\begin{figure}
\includegraphics[width=5 in]{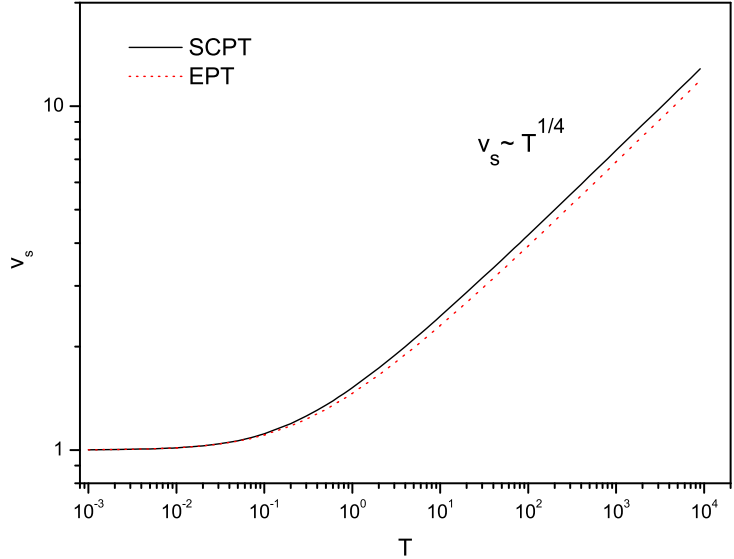}
\caption{\label{fcsT} Effective sound speed $v_s=\sqrt{f}$ as a
function of temperature. At high temperature regime, $v_s\propto
T^{1/4}$. Here $K=1$, $\lambda=1$.}
\end{figure}

\begin{figure}
\includegraphics[width=5 in]{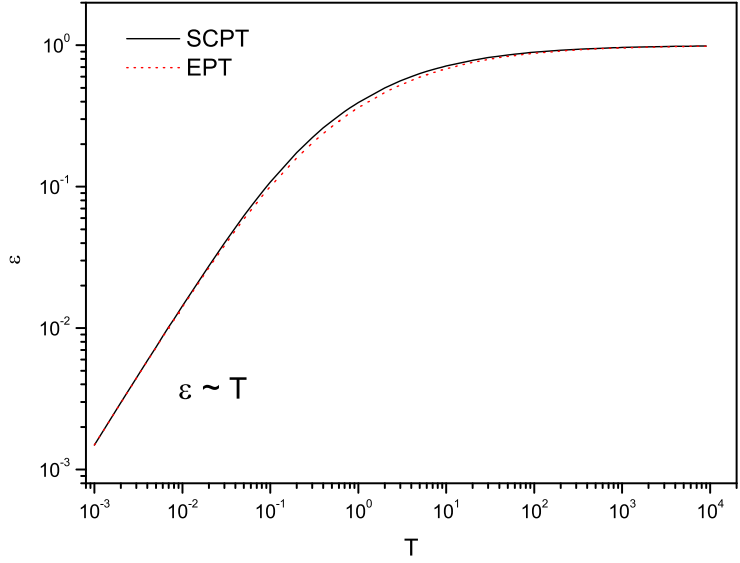}
\caption{\label{fepsT}Temperature dependence of the dimensionless
nonlinearity $\epsilon$. At low temperature regime $\epsilon
\propto T$, while at high temperature limit $\epsilon \simeq 1$.
Here $K=1$, $\lambda=1$.}
\end{figure}

\begin{figure}
\includegraphics[width=5 in]{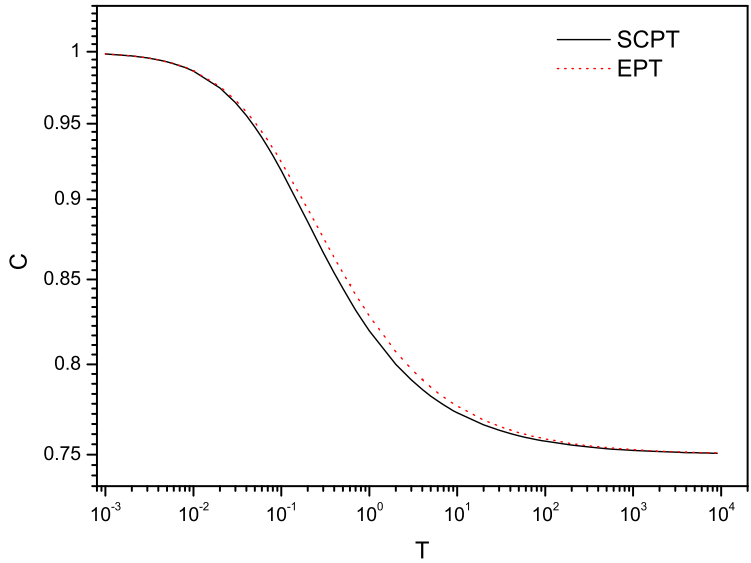}
\caption{\label{CT}Temperature dependence of specific heat. $C
\simeq 1$ at low temperature regime (harmonic limit) and $C$
approaches the lower limit $3/4$ at high temperature regime. Here
$K=1$, $\lambda=1$.}
\end{figure}

In Fig.~\ref{fcsT}, we plot the effective sound speed
$v_s\equiv\sqrt{f}$ (and $\tilde{v}_s\equiv\sqrt{\alpha}$) as a
function of the temperature. The high temperature behavior gives
$v_s\propto T^{1/4}$, which was already reported in
Ref.~\cite{Aoki01, Li_EPT07a}. We also compare $\epsilon$ and $C$
calculated from SCPT and EPT in Fig.~\ref{fepsT} and
Fig.~\ref{CT}, respectively. One can notice that SCPT and EPT
yield practically the same temperature dependence over seven
orders of magnitude.
% [to give some comments/reasons here.]

We will now apply these two effective phonon approaches to
calculate the thermal conductivity from the Debye
formula~\eqref{Debye}. For the sake of simplicity, we will only present the
derivation from SCPT. The phonon group velocity within SCPT is given by
\begin{equation}\label{velocity}
v_k=\frac{\partial \omega_k}{\partial
k}=\sqrt{f}\cos{\frac{k}{2}}\propto \sqrt{f}.
\end{equation}
According to the assumption for the relaxation time~\eqref{tau},
the mean free path reads as
\begin{equation}\label{MFP}
l_k=v_k\tau_k \propto \epsilon^{-1}.
\end{equation}
Substituting Eq.~\eqref{velocity} and Eq.~\eqref{MFP} into
Eq.~\eqref{Debye}, the temperature dependence of the thermal
conductivity can be given by~\cite{Li_EPT07a}
\begin{equation}\label{kappa}
\kappa(T)\propto \frac{C\sqrt{f}}{\epsilon}.
\end{equation}
The calculation of $\kappa$ from EPT follows along similar lines.
By replacing in Eq. (\ref{kappa}) the specific heat, the group
non-linearity and the non-linearity by their counterpart deduced
from EPT, that is Eq.~\eqref{alpha},~\eqref{epsEPT} and
\eqref{SHEPT}, one obtains an alternative expression for the
thermal conductivity (\ref{kappa}). Note that we only consider the
temperature dependence of $\kappa$ here as in
Ref.~\cite{Li_EPT07a}. Nevertheless, it should be emphasized that
$\kappa$ is both temperature and size dependent. $\kappa$ diverges
in the thermodynamic limit due to the divergence of summation with
respect to $k$ in the acoustic regime, which is not the concern of
this study.

\begin{figure}
\includegraphics[width=5 in]{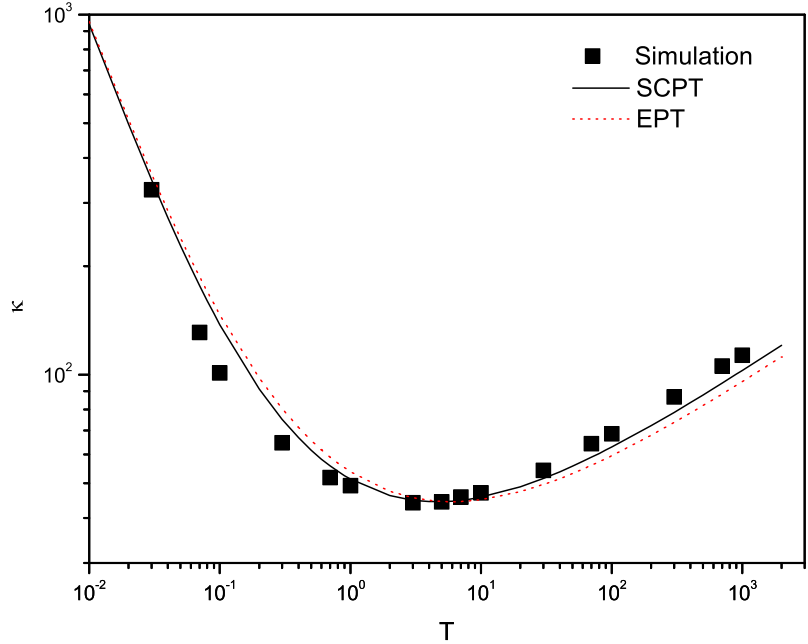}
\caption{\label{KTb1} Temperature dependence of thermal
conductivity. Here $K=1$, $\lambda=1$, and system size $N=1024$
for simulation. Analytical results (solid line and dot line) are
rescaled (divided by a constant), which is consistent with
simulation result in whole temperature range. The simulation
result shows the temperature for the minimum of $\kappa$ at
$T_0\approx 5$, in agreement with Eq.~\eqref{T0}.}
\end{figure}

\begin{figure}
\includegraphics[width=5 in]{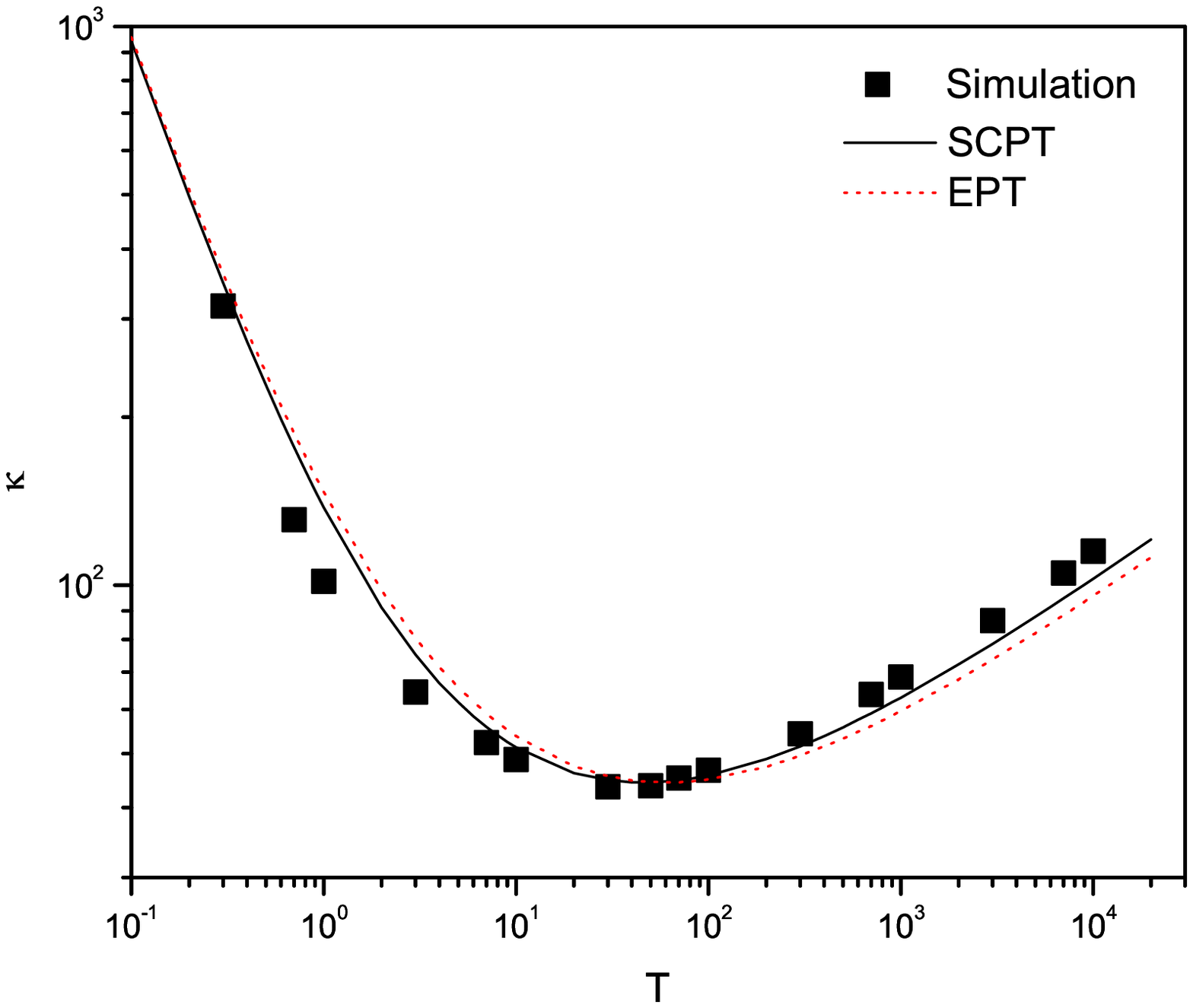}
\caption{\label{KTb01}Temperature dependence of thermal
conductivity. Here $K=1$, $\lambda=0.1$, and $N=1024$ for
simulation. Like Fig.~\ref{KTb1}, analytical results are rescaled.
The simulation result shows $T_0\approx 50$, in agreement with
Eq.~\eqref{T0}.}
\end{figure}

The analytical prediction and the simulation result for the
temperature dependence of $\kappa$ are compared in Fig.~\ref{KTb1}
and Fig.~\ref{KTb01} for $\lambda=1$ and $\lambda=0.1$,
respectively. One can notice that the power law behavior observed
in Ref.~\cite{Aoki01,Li_EPT07a}, i.e., $\kappa \propto T^{-1}$ at
low temperature regime and $\kappa \propto T^{1/4}$ at high
temperature regime are exactly reproduced. As a comparison,
non-equilibrium molecular dynamics simulations were performed by
applying Langevin heat baths at the two ends of the
chain~\cite{Lepri_rep}. In order to compute the thermal
conductivity, a finite temperature difference with $10\%$
deviation from the average temperature was consistently used. It
is clearly seen that the analytical calculations, rescaled by a
constant factor that is implicit in the Debye formula, give good
agreements with the simulation results in a large range of
temperature.


Both Fig.~\ref{KTb1} and Fig.~\ref{KTb01} show the existence of a
minimum of $\kappa$ at a specific temperature $T_0$. By computing the
derivative of the thermal conductivity with respect to $T$,
\begin{equation}
\frac{\partial \kappa}{\partial T}\bigg\vert_{T_0} =0,
\end{equation}
one easily obtains the turning point in the simple form
\begin{equation}\label{T0}
T_0=g_0\frac{K^2}{\lambda}.
\end{equation}
Here we neglect the temperature dependence of the classical
specific heat $C$, which has a very weak dependence on the
temperature. The dimensionless constant $g_0$ is then given by
\begin{equation}\label{g0}
g_0=\frac{7}{3}+\sqrt{5},
\end{equation}
If one takes into account the specific heat given by
\eqref{SHSCPT}, the behavior of $T_0\propto K^2/\lambda$ survives
while $g_0$ should be replaced approximately by $5.52$. The
turning point $T_0$ denotes the transition of $\kappa$ from
decreasing behavior to increment behavior with increasing
temperature.

\begin{figure}
\includegraphics[width=5 in]{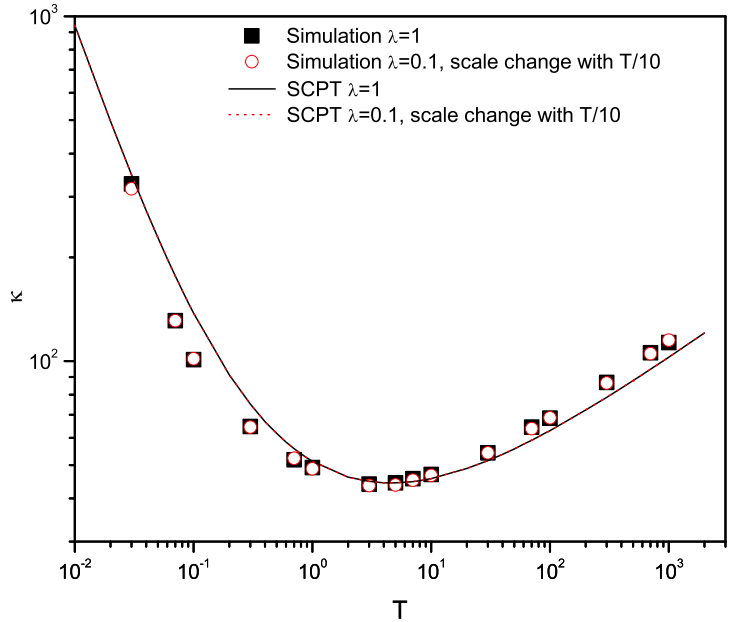}
\caption{\label{scaling}Scaling of thermal conductivity. Here we
simply apply coordinate transformation $\check{T}=T/10$ to
Fig.~\ref{KTb01} and then merge it with Fig.~\ref{KTb1}. One can see
the overlap of the two figures, which indicates the scaling of
thermal conductivity.}
\end{figure}

Eq.~\eqref{T0} can be understood by scaling the
Hamiltonian~\eqref{Ham} as done in Ref.~\cite{Aoki01}. Let
\begin{subequations}\label{x_scaling}
\begin{align}
p_i&=\frac{K}{\sqrt{\lambda}}\check{p}_i,\\
x_i&=\sqrt{\frac{K}{\lambda}}\check{x}_i,
\end{align}
\end{subequations}
one obtains
\begin{equation}\label{En_scaling}
H=\frac{K^2}{\lambda}\check{H},
\end{equation}
where the dimensionless Hamiltonian $\check{H}$ reads as
\begin{equation}
\check{H}=\sum \frac{\check{p}_i^2}{2}
+\frac{1}{2}(\check{x}_{i+1}-\check{x}_i)^2
+\frac{1}{4}(\check{x}_{i+1}-\check{x}_i)^4.
\end{equation}
Eq.~\eqref{x_scaling} leads to
\begin{equation}\label{T_scaling}
T=\langle p_i^2\rangle =\frac{K^2}{\lambda}\langle \check{
p}_i^2\rangle=\frac{K^2}{\lambda}\check{T}.
\end{equation}
Eq.~\eqref{T_scaling} means that the temperature dependence of
physical quantities of the FPU-$\beta$ model can be rescaled according to
the temperature transformation in Eq.~\eqref{T_scaling}, which
can be verified in Fig.~\ref{scaling}.

Note that the nonlinearity $\lambda$ dependence of $\kappa$ is
similar with the temperature dependence shown in Fig.~\ref{KTb1},
Fig.~\ref{KTb01} and Fig.~\ref{scaling}. A critical nonlinearity
$\lambda_0$ corresponding to the minimum of $\kappa$ can also be
similarly obtained as
\begin{equation}\label{lambda0}
\lambda_0=g_0\frac{K^2}{T}.
\end{equation}
Eq.~\eqref{T0} and Eq.~\eqref{lambda0} shows that the temperature
$T$ and the nonlinearity $\lambda$ are inversely equivalent.

Finally, it is interesting to note that
\begin{equation}
\epsilon(T_0)=\epsilon(\lambda_0)=\frac{\sqrt5-1}{2},
\end{equation}
which is the well-known golden ratio. Due to the particular
property of the golden ratio, the ratio of nonlinear potential and
harmonic potential $\langle E_l\rangle/\langle E_n\rangle$ is also
the golden ratio.

\section{Quantum corrections}
From the experimental point of view, the thermal conductivity of real
materials should vanish at zero temperature. This behaviour is
inexistent within the classical description of heat conduction as
shown above, since the latter leads to a divergent thermal
conductivity at zero temperature. The disagreement between the
classical description and experimental curves results from the
inability of the classical physics to take into account the freezing
of phonon modes, which is a pure quantum effect. We would like to
emphasize that the consideration of the low temperature regime is not
of pure theoretical interest. Generally, the Debye temperature of
many solid systems is indeed well above ambient temperature. To name
but a few, $T_D=783$ K in stoichiometric CaB$_6$ and $T_D= 960$ K for
vacancy doped EuB$_6$~\cite{Gianno03}. In carbon nanotubes, the Debye
temperature may even exceed $1000-2000$ K~\cite{Cahill03}. One thus
should take into account quantum effects, namely the partial
thermalization of phonon modes even far above the room temperature
for these materials.

In the following, we will treat the Debye equation for the
FPU-$\beta$ model semi-classically and show that the vanishing
behaviour of $\kappa$ manifests itself if one correctly takes into
account the Bose-Einstein statistics. At this point, we should
recall that the derivation of EPT is based on the classical
equipartition theorem (see Eq.~\eqref{equip}). It is thus
impossible to extent the effective phonon approach to the quantum
regime, which makes SCPT more adequate since the latter can be
derived from a purely quantum approach. The derivation of the
quantum self-consistent phonon theory (QSCPT) is presented in the
appendix. To avoid the divergence of Eq.~\eqref{Debye} due to the
goldstone mode ($k=0$), a small quadratic on-site potential
$f_0x_i^2/2$ is included in the Hamiltonian~\eqref{Ham}. Here the
constant $f_0$ is fixed to its lower-boundary ($f_0=10^{-6}$) so
that its decrease below this limit doesn't change the value of
$\kappa$. A strong point of QSCPT is that the approach
unambiguously provides a mode decomposition for the specific heat
and the phonon velocity, which allows one to consider in a
rigorous way the discrete mode summation of the Debye
equation~\eqref{Debye}. The specific heat per mode $C_k$ is
obtained from~\eqref{ACp}. The phonon velocity $v_k$ follows from
the derivative of the pseudo-phonon spectrum \eqref{A19} and
according to Eq.~\eqref{MFP}, the mean free path is given by
$l_k=\tau_k v_k$ where $\tau_k^{-1}\propto \epsilon\omega_k$.  For
this model, the phonon frequency that should be numerically solved
with Eq. \eqref{A17} is

\begin{equation}\label{SCPeqI}
\omega_p^2=f_0+4\left(K+3\lambda\left<\delta
x^2\right>\right)\sin^2\left(\frac{p\pi}{N}\right),
\end{equation}
where the integer $p\in[1,N]$ and the phonon mode $k\equiv
{2p\pi}/{N}$. The non-linearity parameter is given by
\begin{equation}\label{nonlin3}
\epsilon=\frac{\lambda\left<\delta
x^4\right>/4}{f_0\left<x^2\right>/2+K\left<\delta
x^2\right>/2+\lambda \left<\delta x^4\right>/4}.
\end{equation}

\begin{figure}
\includegraphics[width=5 in]{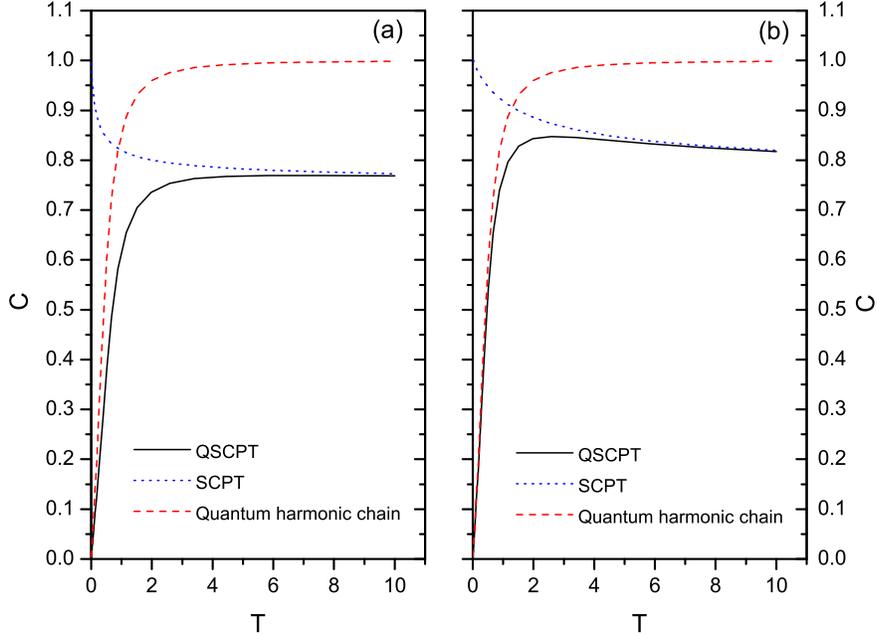}
\caption{\label{QCT}Specific heat as a function of temperature for
(a) $\lambda=1$; (b) $\lambda=0.1$. The harmonic potential is given
by $V_{0}=K\delta x^2/2+f_0x^2/2$. $K=1$ and $f_0=10^{-7}$ for all
cases. At low temperature regime, $C\propto T$.}
\end{figure}

Fig.~\ref{QCT} gives the specific heat $C=1/N\sum C_p$ as a
function of the temperature. At the low temperature regime, $C$ increases
linearly with increasing temperature since the high energy
modes are gradually thermalized. This is the well-known partial thermalization effect.

The agreement with the rigorous result for the harmonic chain
($\lambda=0$) at low temperature regime shows that the harmonic
quantum fluctuations play the dominant role. The quantum effect,
as expected, is negligible above the Debye temperature $T_D=2\sqrt{K}=2$ and the
system behaves like the classical one.

\begin{figure}
\includegraphics[width=5 in]{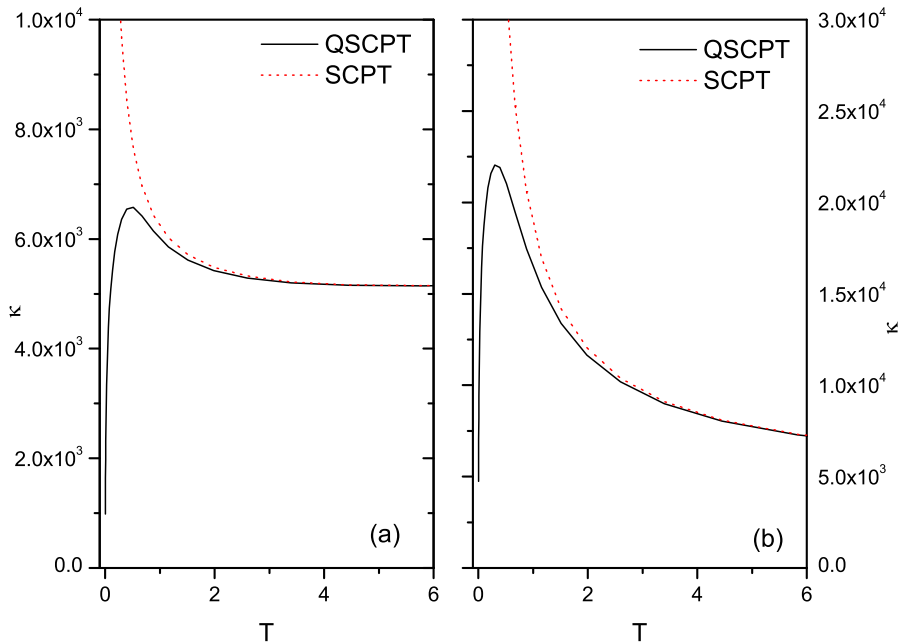}
\caption{\label{QKT}Temperature dependence of thermal conductivity
for (a) $\lambda=1$; (b) $\lambda=0.1$. $K=1$ and $f_0=10^{-6}$ for
both cases.}
\end{figure}

By collecting all the quantities in the Debye
equation~\eqref{Debye}, we finally obtain the temperature
dependence of the thermal conductivity from the quantum approach,
which we compare in Fig.~\ref{QKT} with its classical counterpart.
When the temperature is high enough, QSCPT reproduces the
classical behavior as shown in Fig.~\ref{KTb1} and
Fig.~\ref{KTb01}, where the mean free path and the specific heat
are nearly constant and the temperature dependence of $\kappa$
mainly follows that of the sound speed $v_s\propto T^{1/4}$. The
classical and quantum results agree up to the Debye temperature.
The deviation of the quantum result takes place around $T_D$,
which is followed by a peak at temperature $T_{max}\approx1$, then
the thermal conductivity drops to zero. $T_{max}$ here indicates
the lower bound of the temperature range over which
\textit{umklapp} processes~\cite{Peierls55} yield the dominant
contribution to thermal conductivity. The existence of this peak
results from a competition between the increasing behavior of the
mean free path and the strongly decreasing tendency of the
specific heat. The latter property is a consequence of the partial
thermalization of high frequency phonons, which is a pure quantum
effect explaining the failure of classical approaches below the
Debye temperature. At low temperatures, the mean free path and the
sound speed are nearly constant in the quantum case. The
difference for the temperature dependence of $\kappa$ is mainly
due to the difference of the specific heat. Our result based on
QSCPT, characterized by the existence of the \textit{umklapp}
peak, is qualitatively consistent with the experimental
studies~(see, e.g., \cite{Hone99, Dress00, Cahill02, Cahill03}).
Finally, one should note that the scaling
behavior~\eqref{T_scaling} is not applicable for quantum case.

\section{Summary}
In summary, SCPT and EPT are compared in the present study. The
basic difference between SCPT and EPT lies in the different way to
calculate the statistical average of physical quantities.
Specifically, the canonical average is applied in EPT but Gaussian
average is applied in SCPT. We show that both SCPT and EPT give the
same behavior for the thermodynamic quantities of a classical
system, which indicates their equivalence. It seems that gaussian
average is preferable for analytical derivation, especially for
models with an on-site potential, since the thermal averages should
be computed in this case from numerical transfer matrix
calculations. The additional gain is that SCPT can be extended to
study the quantum system, as shown in section II. Note that EPT is
based on the assumption of the equipartition theorem, whose validity
in the quantum regime is broken. Thus QSCPT offers a simple way to
explore for a given Hamiltonian model the low temperature physics
related to the Bose-Einstein statistics.

We studied the temperature dependence of the thermal conductivity of
the FPU-$\beta$ model for both the classical and quantum case. The
temperature for the minimum of $\kappa (T)$ was determined from the
classical SCPT, which clearly shows the scaling behavior of thermal
conductivity for this non-linear model. We also showed that
non-equilibrium molecular dynamics simulations are in good agreement
with the analytical results. At the low temperature regime, the
semiclassical treatment of the Debye equation reproduced the
\textit{umklapp} peak, a well-known characteristic observed in
experimental studies.

\begin{acknowledgments}
This work was supported in part by grants from the Hong Kong
Research Grants Council (RGC) and the Hong Kong Baptist
University.
\end{acknowledgments}

\appendix*
\section{Quantum Self-Consistent Phonon Theory}

In this appendix, we present a simple derivation of the first
order self-consistent phonon theory for 1D Hamiltonian systems.
The presented approach is similar to the derivation of the
variational path integral method (see~\cite{Feynman86, Kleinert95,
Giachetti86}). In the path integral representation of quantum
statistical mechanics, the $N$-body partition function in the
canonical ensemble can be expressed as a path integral over
periodic trajectories, that is
\begin{equation}\label{A1}
Z=\int\mathbf{\emph{D}x}\hspace{0.5mm}e^{-S[\mathbf{x}]},
\end{equation}
where
\begin{equation}\label{A2}
S=\int_0^{\hbar\beta}
d\tau\left(\frac{m}{2}\dot{\mathbf{x}}^2+U[\mathbf{x}]\right)
\end{equation}
is the Euclidean action and the first and second terms in the
integral correspond to the total kinetic and potential energies.
It is well known that an exact evaluation of the $N$-body path
integral (\ref{A1}) for general anharmonic potentials is
impossible. The idea of SCPT is basically approximating the
original Euclidean action with a trial action that allows an exact
evaluation of the trace (\ref{A1}). Since we deal in this work
exclusively with 1D Hamiltonian models with closest neighbor
interactions of the form
\begin{equation}\label{A3.5}
H=\sum_{k=1}^N\left\{\frac{m}{2}\dot{x}_k^2+ V(x_k)+W(\delta
x_k)\right\},
\end{equation}
where $\delta x_k= x_k-x_{k-1}$,  an optimal choice of the trial
Hamiltonian is that of a coupled harmonic oscillator chain,
\begin{equation}\label{A3}
H_0=\sum_{k=1}^N\left\{\frac{m}{2}\dot{x}_k^2
+\frac{\lambda_1}{2}x_k^2+\lambda_2x_k+g x_k x_{k-1}\right\}.
\end{equation}
The trial parameters $\lambda_1$, $\lambda_2$ and $g$ are to be
deduced by minimizing the right hand-side of the Feynman-Jensen
inequality,
\begin{equation}\label{A4}
\mathcal{F}\leq \mathcal{F}_0 +\left<H-H_0\right>,
\end{equation}
where $\mathcal{F}_0=-k_BT\ln Z_0$. The trial partition function
given by
\begin{equation}\label{A5}
Z_0=\int\mathbf{\emph{D}x}\hspace{0.5mm}e^{-S_0[\mathbf{x}]/\hbar},
\end{equation}
can be easily computed by first performing a Fourier decomposition
of the periodic paths,
\begin{equation}\label{A6}
x_k=\sum_{n\geq0}\left(x_{kn}e^{i\Omega_nt}+c.c.\right),
\end{equation}
where $\Omega_n=2\pi n/(\beta\hbar)$ stand for Matsubara
frequencies and the measure of integration of Eq. (\ref{A5}) has
the form
\begin{equation}\label{A7}
\mathbf{\emph{D}x}=\prod_{k,n}\frac{dx_{kn}^{re}\hspace{0.5mm}dx_{kn}^{im}}{\sqrt{2\pi\beta\hbar^2/m}}.
\end{equation}
In the last expression as well as in Eq. (\ref{A6}), index $k$ run
over oscillators and $n$ over Fourier components. We then
substitute the expansion (\ref{A6}) into (\ref{A3}) and obtain the
trial action by performing the integration over time $\tau$,
\begin{equation}\label{A8}
S_0=\int_0^{\beta\hbar}d\tau H_0.
\end{equation}
The trial action $S_0$ that follows is a quadratic function of the
Fourier components $x_{kn}$, that is
\begin{eqnarray}
S_0&=&\beta\hbar\sum_{k=1}^N\left\{\frac{\lambda_1}{2}x_{k0}^2+\lambda_2x_k+g x_{k0} x_{k-10}\right\}\nonumber\\
&+&\beta\hbar\sum_{k=1}^N\sum_{n\geq1}^N\left\{(\lambda_1+m\Omega_n^2)\left(\left|x_{kn}^{re}\right|^2
+\left|x_{kn}^{im}\right|^2
\right)+2g\left(x_{kn}^{re}x_{k-1n}^{re}+x_{kn}^{im}x_{k-1n}^{im}\right)\right\}\label{A8.2}
\end{eqnarray}

The next step consists of the trivial integration in Eq.
(\ref{A5}) over Fourier components, which yields
\begin{equation}\label{A9}
Z_0=\prod_{p=1}^N\frac{\sin(p\pi/N)}{\sinh(\beta\hbar\omega_p/2)},
\end{equation}
for $\lambda_1+2g=0$ and
\begin{equation}\label{A10}
Z_0=e^{N\beta\frac{\lambda_1+2g}{2}\eta^2}\prod_{p=1}^N\frac{1}{2\sinh(\beta\hbar\omega_p/2)},
\end{equation}
otherwise. In the last two expressions which give the zero-th
order free energy $\mathcal{F}_0$, we have defined the
pseudo-phonon frequencies in the form
\begin{equation}\label{A11}
m\omega_p^2=\lambda_1+2g-4g\sin\left(\frac{p\pi}{N}\right)
\end{equation}
and an additional parameter
\begin{equation}\label{A12}
\eta=-\frac{\lambda_2}{\lambda_1+2g}.
\end{equation}

The calculation of the first order correction to free energy
\begin{equation}\label{A13}
\left<H-H_0\right>=\frac{1}{Z_0}\int\mathbf{\emph{D}x}\hspace{0.5mm}e^{-S[\mathbf{x}]}\left\{\sum_{k=1}^N
\left[V(x_k)+W(\delta x_k)\right]-H_0\right\}
\end{equation}
proceeds in a similar way. The usual trick consists in expanding
the on-site and inter-site potential in Fourier basis,
\begin{equation}\label{A14}
V(x_k)=\int\frac{dq}{2\pi}\tilde{V}(q)e^{iqx_k},\hspace{1cm}W(\delta
x_k)=\int\frac{dq}{2\pi}\tilde{W}(q)e^{iq \delta x_k}.
\end{equation}
Then we evaluate the average potentials per particle
$\left<V(x_k)\right>$ and $\left<V(x_k)\right>$ in (\ref{A13}) by
integrating over Fourier components and finally invert the fourier
transforms of Eq. (\ref{A14}). Consequently, the average value of
the potential energies can be expressed in the form of smeared-out
potentials,
\begin{equation}\label{A15}
V_\rho(\eta)\equiv\left<V(x_k)\right>
=\int\frac{dy}{\sqrt{2\pi\rho^2}}e^{-\frac{(y-\eta)^2}{2\rho^2}}V(y),
\hspace{1cm} W_\gamma\equiv\left<V(\delta
x_k)\right>=\int\frac{dy}{\sqrt{2\pi\gamma^2}}e^{-\frac{y^2}{2\gamma^2}}W(y),
\end{equation}
where we have defined two parameters $\rho^2$ and $\gamma^2$ which
correspond to lattice displacement and two-point correlation
function, that is
\begin{equation}\label{A16}
\rho^2\equiv\left<x_k^2\right>
=\frac{\hbar}{2Nm}\sum_p\omega_p^{-1}\coth\left(\frac{\beta\hbar\omega_p}{2}\right)
\end{equation}
and
\begin{equation}\label{A17}
\gamma^2\equiv\left<(x_k-x_{k-1})^2\right>
=\frac{\hbar}{2Nm}\sum_p\frac{4\sin^2\left(\frac{p\pi}{N}\right)}
{\omega_p}\coth\left(\frac{\beta\hbar\omega_p}{2}\right).
\end{equation}
The gaussian smearing (\ref{A15}) is a key characteristic of
variational methods.

The first order free energy per particle can be finally expressed
as
\begin{equation}\label{A18}
F_1=F_0-\sum_p\left\{
\frac{\beta\hbar\omega_p}{4}\coth\left(\frac{\beta\hbar\omega_p}{2}\right)\right\}
+V_\rho(\eta)+W_\gamma.
\end{equation}
By minimizing this expression with respect to $\omega_p$, one
obtains the optimal pseudo-phonon frequency
\begin{equation}\label{A19}
\omega_p^2=\frac{2}{m}\left\{\frac{\partial V_\rho}{\partial
\rho^2}+4\sin^2\left(\frac{p\pi}{N}\right)\frac{\partial
W_\gamma}{\partial \gamma^2}\right\}.
\end{equation}
On the other hand, the variation of Eq. (\ref{A18}) with respects
to $\eta$ yields
\begin{equation}\label{A20}
\frac{\partial V_\rho}{\partial \eta}=0.
\end{equation}
Note that $\eta=0$ for pair on site potentials ($V(x)=V(-x)$).
Using (\ref{A15}), (\ref{A16}) and (\ref{A19}) in Eq. (\ref{A18}),
we can further simplify the first order free energy in the form
\begin{equation}\label{A21}
F_1=F_0 +V_\rho-\rho^2\frac{\partial V_\rho}{\partial
\rho^2}+W_\gamma-\gamma^2\frac{\partial W_\gamma}{\partial
\gamma^2}.
\end{equation}
The specific heat per mode can be then defined by
\begin{equation}\label{ACp}
C_p=-T\frac{\partial^2F_1}{\partial T^2}.
\end{equation}

One can easily obtain the classical limit of~\eqref{A21} by
letting $\hbar\rightarrow0$. The final result is
\begin{equation}\label{A22}
F_1=\frac{1}{N\beta}\sum_p\ln\left(\beta\omega_p\right)
+V_{\rho_c}-\rho_c^2\frac{\partial V_{\rho_c}}{\partial
\rho_c^2}+W_{\gamma_c}-\gamma_c^2\frac{\partial
W_{\gamma_c}}{\partial \gamma_c^2},
\end{equation}
where $\rho^2$ and $\gamma^2$ in Eq. (\ref{A21}) are replaced by
their classical counterpart,
\begin{equation}\label{A23}
\rho_c^2\equiv\left<x_k^2\right>_{\hbar\rightarrow
0}=\frac{1}{Nm\beta}\sum_p\omega_p^{-2}
\end{equation}
and
\begin{equation}\label{gammaC}
\gamma_c^2\equiv\left<(x_k-x_{k-1})^2\right>_{\hbar\rightarrow
0}=\frac{1}{Nm\beta}\sum_p\frac{4\sin^2\left(\frac{p\pi}{N}\right)}{\omega_p^2}.
\end{equation}

For FPU-$\beta$ model~\eqref{Ham},
\begin{equation}\label{A27}
W(\delta x_i)=\frac{K}{2}\delta x_i^2+\frac{\lambda}{4}\delta
x_i^4, \hspace{1cm} V=0.
\end{equation}
Solving Eq.~\eqref{gammaC} self-consistently with Eq.~\eqref{A19},
it is easy to get the effective harmonic constant
\begin{equation}\label{f}
f=K+3\lambda\gamma_c^2 =\frac{K+ \sqrt{K^2+12\lambda K_BT}}{2}.
\end{equation}

%----------------  Reference ---------------------------%
%\bibliography{References}% Produces the bibliography via BibTeX.

\end{document}